\documentclass[12pt]{article}
\usepackage[english]{babel}
\usepackage[utf8]{inputenc}
\usepackage[T1]{fontenc}

\usepackage{amsmath}
\def\bC{\mathbf{C}}

\def\mC{\mathcal{C}}
\usepackage{amsfonts}

\usepackage{url}
\usepackage[dvips]{graphicx}
\usepackage{calc}
\usepackage{subfigure}
\usepackage{multirow}

\def\mH{\mathcal{H}}
\def\mD{\mathcal{D}}

\def\bx{\mathbf{x}}
\def\tmH{\tilde{\mH}}
\newcommand{\pb}[1]{\left\{#1\right\}}

\def\bA{\mathbf{A}}

\def\mR{\mathcal{R}}

\def\mG{\mathcal{G}}
\def\mS{\mathcal{S}}
\def\bG{\mathbf{G}}
\def\ba{\mathbf{a}}
\def\by{\mathbf{y}}

\def\mM{\mathcal{M}}
\def\tmH{\tilde{\mH}}
\def\bH{\mathbf{H}}
\def\bM{\mathbf{M}}

\begin{document}
	\begin{titlepage}
	\begin{center}
		{\Large{ \bf  Deparametrization of General Relativity by Space-Time Filling Unstable D9-Brane with Arbitrary Value of Tachyon	}}
		
		\vspace{1em}  
		
		\vspace{1em} J. Kluso\v{n} 			
		\footnote{Email addresses:
			klu@physics.muni.cz  }\\
		\vspace{1em}
		\textit{Department of Theoretical Physics and
			Astrophysics, Faculty of Science,\\
			Masaryk University, Kotl\'a\v{r}sk\'a 2, 611 37, Brno, Czech Republic}
		
		\vskip 0.8cm
		
		%
		%
		%
		%
		%
		%
		
		\vskip 0.8cm
		
	\end{center}

\begin{abstract}
	 We calculate algebra of constraints
	of deparametrized General Relativity with   space-time filling
	unstable D-brane   for arbitrary
	value of tachyon field $T$. We also propose observables that 
	have vanishing Poisson brackets with all first class constraints.
	\end{abstract}

\bigskip

\end{titlepage}

\newpage

\section{Introduction and Summary}\label{first}
In our previous paper \cite{Kluson:2024rnv} we analyzed space-time filling unstable D9-brane in Type IIA theory 
\cite{Sen:1999md,Garousi:2000tr,Bergshoeff:2000dq,Kluson:2000iy}
in general space-time. 
 We performed deparametrization of theory and calculated algebra of constraints at the asymptotic region of  tachyon field space where 
we showed that such a system is natural for deparametrization of gravity \cite{Brown:1994py}. Deparametrization is general procedure when the original Hamiltonian constraint is 
replaced with following one in the form \cite{Thiemann:2006up} $
\mC=\pi+H\approx 0$ where $\pi$ is momentum conjugate to scalar field $\phi$ and where $H$ is positive function of remaining phase space degrees of freedom which does not depend on $\pi$.
This procedure was used by   T. Thiemann in \cite{Thiemann:2006up} when he suggested possible solution 
of  problem of time in gravity. 

In more details, the problem of time has following origin.
According to Dirac \cite{Dirac} all observables have to be constant along gauge orbits and therefore have vanishing Poisson brackets with all first class   constraints that are generators of gauge transformations. In case of General Relativity it can be shown that the Hamiltonian is given as sum of the first class constraints and hence Hamiltonian vanishes on the constraint surface \cite{Dirac:1958sc,Arnowitt:1962hi}. Then it is clear that all proper observables do not evolve dynamically which is known as problem of time. 
The fact that there is no evolution with respect to the time is in conflict with our everyday experience. On the other hand the canonical Hamiltonian 
describes evolution with respect to coordinate time which really does not have physical meaning thanks to  the manifest
diffeomorphism invariance of any theory coupled to gravity. What really makes sense is evolution with respect
to other fields. 

In our previous  paper \cite{Kluson:2024rnv} we applied procedure suggested in \cite{Thiemann:2006up}
to the case of space-time filling non-BPS D9-brane in Type IIA theory where the tachyon field is used
for deparametrization of  General Relativity  in the regime of  large tachyon field so that tachyon is directly related to the time evolution.  We should stress that the idea that  open string tachyon could be related to the physical time was suggested by A. Sen in  seminal paper \cite{Sen:2002qa} and recently in  \cite{Sen:2023qya} where it was again emphasized that tachyon could be related to physical time when tachyon approaches its vacuum value. It is well known that open string tachyon is special since the vacuum corresponds to the vanishing tachyon potential, for review, see \cite{Sen:2004nf}.

The crucial presumption of the paper \cite{Kluson:2024rnv}  was  that the tachyon is large and hence we can neglect all terms proportional to $V^2$ where $V$ is tachyon potential that appears in the Hamiltonian constraint. In the present work we  relax this condition and study the situation when tachyon  takes any value. We would like to stress that even in this case we can perform deparametrization of gravity however it is not clear whether Poisson bracket between new Hamiltonian constraints is strongly zero or whether it
is proportional to linear combinations of primary constraints.  We explicitly perform calculations of these Poisson brackets and we show that they are really zero \footnote{For
previous work, see \cite{Kuchar:1995xn}.}.

Having derived this important result it is natural to   proceed to the construction of  Dirac variables in the same way as in case \cite{Kluson:2024rnv}, following \cite{Thiemann:2006up}.
However now  we immediately find that such a naive construction cannot work due to the fact that  $H(\bx)$ does not have vanishing Poisson bracket with Hamiltonian constraint thanks to the explicit dependence of $H$ on $T$. In other words the case of finite $T$ needs more general treatment which is based on relational observables, for review see 
\cite{Tambornino:2011vg,Giesel:2017roz}. We try to implement this procedure in our case  following  seminal papers \cite{Dittrich:2004cb,Dittrich:2005kc} but it turns out  that this general mechanism cannot be applied directly to the case of the system of unstable D9-brane coupled to gravity from following reason. For construction of Dirac variables \cite{Dittrich:2004cb,Dittrich:2005kc}
it is necessary to have the same number of clock fields as the number of first class constraints which is ten in case of space-time filling non-BPS D9-brane. On the other hand there is only one scalar field on the world-volume of space-time filling unstable D9-brane, which is tachyon, that can serve as "clock" for Hamiltonian constraint while there are no clock fields for nine spatial diffeomorphism constraints. For that reason we propose Dirac observable whose construction  will not be as general as the one developed in \cite{Dittrich:2004cb,Dittrich:2005kc}.  Explicitly, we construct Dirac observable using the  Hamiltonian constraint $\mG=\frac{1}{\sqrt{\lambda}}p_T+H$. We will also presume that the parameter that appears in Dirac observable is constant over the whole spatial section 
in the same way as in \cite{Thiemann:2006up}. Finally we define
this observable with partial observable that is also invariant
under spatial diffeomorphism which is again crucial presumption. 
Then we will show that this observable strongly Poisson commutes
with smeared form of Hamiltonian constraint. We also show that
it Poisson commutes with spatial diffeomorphism constraint and we determine evolution equation for her. Now due to the fact
that Hamiltonian constraint explicitly depends on $T$ for finite $T$ this evolution equation cannot have the form of  Poisson bracket between this Dirac observable and some Hamiltonian function. We also show that the right side of evolution equation has the form of Poisson bracket between 
this Dirac observable and Hamiltonian in  in the asymptotic region $T\rightarrow \infty$ in agreement with  \cite{Thiemann:2006up}.

Let us outline our results and suggest possible extension of this work. We study deparametrized theory for space-time filling unstable D9-brane coupled to gravity for arbitrary value of tachyon. As the main result we explicitly check that the Poisson bracket between deparametrized Hamiltonian constraints is zero even in this general case. We mean that this is really remarkable result which is especially important when we construct Dirac observable. We argue that this observable has to  be defined using Hamiltonian constraint instead of the Hamiltonian function that was used in case of asymptotic large value of tachyon field. Then we show that this observable has vanishing Poisson brackets with all first class constraints and hence it is true Dirac observable. 

The present analysis suggests that open string tachyon has natural interpretation as time variable even in case of its finite value. Then it would be interesting to study cosmological consequences of this model. We hope to return to this problem in future. 

This paper is organized as follows. In the next section (\ref{second}) we review basic facts about space-time filling
unstable D9-brane and determine its Hamiltonian formulation in deparametrized form. In section (\ref{third}) we calculate
Poisson brackets between smeared form of Hamiltonian constraints and we show that they strongly vanish. In section 
(\ref{third}) we define observables as in case of the vanishing tachyon field and we show that this procedure does not lead to Dirac observables. Finally in section (\ref{fourth}) we 
introduce alternative form of the extended variable and we show that it is true Dirac observable. 

\section{Review of the Basic Facts About Space-Time  Filling Non-BPS D9-Brane}\label{second}
In this section we review basic facts about space-time 
filling  non-BPS D9-brane coupled to gravity. Recall that the
action for this system has the form
\begin{equation}
	S=S_{GR}+S_{nonD} \ , 
\end{equation}
where
\begin{equation}
S_{GR}=\frac{1}{\kappa}\int d^{10}x \sqrt{-g}R(g)
\end{equation}
and where 
\begin{equation}\label{Snongen}
	S_{nonD}=-\int d^{10}x V(T)\sqrt{-\det \bA_{MN}} \ , 
\end{equation}
where
\begin{equation}
	\bA_{MN}=g_{MN}+
	\lambda F_{MN}+\lambda\partial_M T\partial_N T \ , 
\end{equation}
where $\lambda=2\pi\alpha'=l_s^2$, where $l_s$ is string length,  and $F_{MN}=\partial_M A_N-\partial_N A_M$, where 
$A_M$ is a gauge field living on the world-volume of D9-brane,  $T$ is the tachyon field with the potential $V(T)$ with the property that $T$ has two stable
minima for $T_{min\pm}=\pm \infty$ where $V(T_{min \pm})=0$ while it has an unstable maximum at $T_{max}=0$ where $V(T_{max})=\tau_{nonBPS}$ where $\tau_{nonBPS}=\frac{\sqrt{2}2\pi}{(2\pi l_s)^{10}}$ is a tension of unstable Dp-brane. 

In \cite{Kluson:2024rnv} we derived canonical form of the action for this system that has the form 
\begin{eqnarray}
&&	S=S_{GR}+S_{nonD}=\nonumber \\
&&=\int d^{10}x (\pi^{ij}\partial_0 h_{ij}+
p_T\partial_0 T+\pi^i \partial_0 A_i-N\mC-N^i\mC_i+A^0\mathcal{E}) \ , \nonumber \\
\end{eqnarray}
where
\begin{eqnarray}
&&\mC=\frac{\kappa}{\sqrt{h}}(\pi^{ij}h_{ik}h_{jl}\pi^{kl}-
(\pi^{ij}h_{ij})^2)-\frac{1}{\kappa}\sqrt{h}r+
\sqrt{\mD+\mH_ih^{ij}\mH_j}\equiv \nonumber \\
&& \equiv \mH^G_\bot+\mC^{matt} \ , \nonumber \\
&&\mD=\frac{1}{\lambda}p_T^2+\frac{1}{\lambda^2}\pi^i\bA^S_{ij}\pi^j+V^2\det \bA_{ij} \ , \nonumber \\
&& \bA_{ij}=h_{ij}+\lambda \partial_iT\partial_j T+
\lambda F_{ij}  \ , 
\bA_{ij}^S=\frac{1}{2}(\bA_{ij}+\bA_{ji})=
h_{ij}+\lambda T\partial_jT \ , 
 \nonumber \\
&&\mC_i=
-2\nabla_l\pi^{kl}h_{ki}+\mH_i \equiv \mH^G_i+\mH_i \equiv \tmH_i+p_T\partial_i T\approx 0 \ , \nonumber \\
&&\mathcal{E}=\partial_i\pi^i \approx 0 \ , \quad 
\mH_i=p_T\partial_iT+F_{ij}\pi^j \ , \quad 
\tmH_i=\mH_i^G+F_{ij}\pi^j \ , \nonumber \\
\end{eqnarray}
and where we  introduced $9+1$ formalism for the background gravity \footnote{For  review, see
	\cite{Gourgoulhon:2007ue}.}. We considered $10-$dimensional manifold
$\mathcal{M}$ with the coordinates $x^M \ , M=0,\dots,9$ and
where $x^M=(t,\bx) \ , \bx=(x^1,x^2,\dots,x^{9})$. This 
space-time is endowed with the metric $g_{MN}(x^\rho)$
with signature $(-,+,\dots,+)$ and it is foliated  by a family of space-like surfaces $\Sigma$ defined by
$t=x^0=\mathrm{const}$.  $h_{ij}, i,j=1,2,\dots,9$ denotes the metric on $\Sigma$
with inverse $h^{ij}$ so that $h_{ij}h^{jk}= \delta_i^k$. We further
introduced the operator $\nabla_i$ that is covariant derivative
defined with the metric $h_{ij}$.
We also defined  the lapse
function $N=1/\sqrt{-g^{00}}$ and the shift functions
$N^i=-g^{0i}/g^{00}$. In terms of these variables 
the components of the metric $g_{MN}$ were written as 
\begin{eqnarray}
	g_{00}=-N^2+N_i h^{ij}N_j \ , \quad g_{0i}=N_i \ , \quad
	g_{ij}=h_{ij} \ ,
	\nonumber \\
	g^{00}=-\frac{1}{N^2} \ , \quad g^{0i}=\frac{N^i}{N^2} \
	, \quad g^{ij}=h^{ij}-\frac{N^i N^j}{N^2} \ .
	\nonumber \\ . 
\end{eqnarray}
We further  also used the fact that requirement of the preservation 
of the constraint $\pi^0\approx 0$ implies the constraint $\mathcal{E}\approx \partial_i\pi^i\approx 0$
\footnote{In can be easily shown that $\mathcal{E}\approx 0$ is first class constraint and it is generator of gauge transformations for the gauge field $A_i$. In what follows we will presume that all observables are invariant under this gauge transformation.}. Note that $\pi^{ij}$ are momenta conjugate to $h_{ij}$ and $r$ is scalar curvature calculated with the metric $h_{ij}$. Finally   the requirement of  preservation of  constraints 
$\pi_N\approx 0 \ , \pi_{N^i}\approx 0$ where $\pi_N$ and $\pi_{N^i}$ are momenta conjugate to $N$ and $N^i$ respectively led to an existence of the constraints 
\begin{equation}
	\mC\approx 0 \ ,  \quad  \mC_i\approx 0 \ . 
\end{equation}	
We  proved  in \cite{Kluson:2024rnv} that it is possible to replace $\mC\approx 0$ with new constraint $\mG\approx 0$ that has the form
\begin{eqnarray}\label{mGdef}
	&&\mG \equiv \frac{1}{\sqrt{\lambda}}p_T+\sqrt{\frac{1}{2}((\mH^{G}_\bot)^2-\mS)
		+\sqrt{\frac{1}{4}(\mS-(\mH^{G}_\bot)^2)^2
			-\mR}}\equiv
	\nonumber \\
	&&	\equiv  \frac{1}{\sqrt{\lambda}}p_T+H \  
	\approx 0 \ , \nonumber \\
\end{eqnarray}
and where $\mS$ and $\mR$ are
defined as
\begin{eqnarray}
	&&\mS\equiv 
\frac{1}{\lambda^2}\pi^i h_{ij}\pi^j
	+V^2\det \ba+\mH_i^{G}h^{ij}\mH_j^{G} \ , \nonumber \\
	&&\mR=\frac{1}{\lambda^2}\pi^i\tmH_i\tmH_j\pi^j+V^2\det \ba\ba^{ij}\tmH_i\tmH_j\approx 0 \ ,
\quad 
 \ba_{ij}=h_{ij}+\lambda F_{ij} \ .
\end{eqnarray}
Note that $\pi^i
\tmH_i=
\pi^i 
\mH_i^G$ due to the fact that $\pi^iF_{ij}\pi^j=0.$
%
Further,  $\ba^{ij}$ is matrix inverse to $\ba_{ij}$
\begin{equation}
\ba_{ij}\ba^{jk}=\delta_i^k \ . 
\end{equation}

One of the goals of this paper is to calculate Poisson brackets between constraints $\mG(\bx)$ and $\mG(\by)$. In our previous paper 
\cite{Kluson:2024rnv} we shown that in the case of asymptotic 
large $T$, when we can neglect terms proportional to $V^2$, the Poisson brackets between constraints $\mG(\bx),\mG(\by)$ vanish strongly. In this paper we would like to extend this analysis to the case of arbitrary value of tachyon. 
 We will explicitly calculate algebra of constraints $\mG(\bx),\mG(\by)$ 
 with the help of canonical Poisson brackets
\begin{eqnarray}
	&&	\pb{h_{ij}(\bx),\pi^{kl}(\by)}=
	\frac{1}{2}(\delta_i^k\delta_j^l+\delta_i^l\delta_j^k)
	\delta(\bx-\by) \ , \quad 
		\pb{T(\bx),p_T(\by)}=\delta(\bx-\by) 
 \ , \nonumber \\
	&&		\pb{A_\alpha(\bx),\pi^\beta(\by)}=\delta_\alpha^\beta\delta(\bx-\by) \ , \quad \alpha,\beta=0,1,\dots,9 	
\ . \nonumber \\	
\end{eqnarray}
First of all we define following object
\begin{equation}
G(\bx)=(H^G_\bot)^2-H_i^G h^{ij}H_j^G \ , \quad 
\bG(X)=\int d^9\bx X(\bx)G(\bx) \ , 
\end{equation}
whose importance was firstly stressed in 
\cite{Brown:1994py} and further studied in \cite{Kuchar:1995xn,Markopoulou:1996nh,Thiemann:2006up}. Note that
it is function of gravitational variables only. 
In our previous paper  we reproduced an important result \cite{Brown:1994py}
\begin{equation}
\pb{\bG(X),\bG(Y)}=0 \ .
\end{equation}
Further, due to the complex form of the expression $H$ we will calculate very carefully  
 Poisson brackets between individual terms that appear in $H$. We start with $\mM$ where $\mM$ is defined as
 \begin{equation}\label{defmM}
 	\mM=(\mH_\bot^G)^2-
 	\mS \ . 
 \end{equation}
 	 It is convenient to introduce smeared form of this expression which again allows us to avoid to  work with partial derivatives of delta functions. Explicitly, let us introduce  $\bM(X)=\int d^9\bx X\mM$ 
and  calculate  
\begin{eqnarray}\label{pbbM}
&&\pb{\bM(X),\bM(Y)}=
\pb{\bG(X),\bG(Y)}-\nonumber \\
&&\pb{\bG(X),\int d^9\by Y(\frac{1}{\lambda^2}\pi^i h_{ij}\pi^j+V^2\det \ba)}
-\nonumber \\
&&-\pb{\int d^9\bx X(\frac{1}{\lambda^2}\pi^i h_{ij}\pi^j+V^2\det \ba),
	\bG(Y)}+
\nonumber \\
&&\pb{\int d^9\bx X(\frac{1}{\lambda^2}\pi^i h_{ij}\pi^j+V^2\det \ba),
\int d^9\by Y (\frac{1}{\lambda^2}\pi^i h_{ij}\pi^j+V^2\det \ba)}
\nonumber \\
&&=4\int d^9\bx(X\partial_iY-Y\partial_iX)\frac{1}{\lambda^2}\pi^i\tmH_j\pi^j +
\nonumber \\ 
&&2\int d^9\bx (X\partial_kY-Y\partial_kX)(\mH^G_l \ba^{lk}+\ba^{kl}\mH_l^G)V^2\det \ba+
\nonumber \\
&&
+2\int d^9\bx (X\partial_iY-Y\partial_iX)(\ba^{ik}+\ba^{ki})F_{km}\pi^m V^2\det \ba \ , \nonumber \\
\end{eqnarray}
%
where we used the fact that 
\begin{eqnarray}
	\pi^i\tmH_i=
\pi^i(\mH_i^G+F_{ij}\pi^j)=
\pi^i\mH_i^G	\ . 
\nonumber \\
\end{eqnarray}
and also 
\begin{eqnarray}
&&\pb{\int d^9\bx X\mH_i^G h^{ij}\mH_j^G,\int d^9\by Y V^2\det \ba}+
\pb{
\int d^9\by X V^2\det \ba,\int d^9\bx Y\mH_i^G h^{ij}\mH_j^G}=\nonumber \\	
&&=-2\int d^9\bx (\partial_kXY-X\partial_k Y)(\mH^G_l \ba^{lk}+\ba^{kl}\mH_l^G)V^2\det \ba \ , 
\nonumber \\
\end{eqnarray}
where we have also used
\begin{equation}
	\pb{\int d^9\bx X^i\mH_i^G,h_{kl}}=-X^m\partial_m h_{kl}-\partial_k X^m h_{ml}-
	h_{km}\partial_l X^m \ . 
\end{equation}
Finally we calculate the last expression in (\ref{pbbM}) 
\begin{eqnarray}
&&	\pb{\int d^9\bx X(\frac{1}{\lambda^2}\pi^i h_{ij}\pi^j+V^2\det \ba),
		\int d^9\by Y (\frac{1}{\lambda^2}\pi^i h_{ij}\pi^j+V^2\det \ba)}=
	\nonumber \\
&&=\frac{2}{\lambda}\int d^9\bx (X\partial_kY-Y\partial_k X) \pi^jh_{ji}(\ba^{ik}-\ba^{ki})V^2\det \ba
\nonumber \\
&&=-2\int d^9\bx (X\partial_k Y-Y\partial_kX)\pi^j F_{ji}(\ba^{ik}+\ba^{ki})V^2\det \ba \ , \nonumber \\
\end{eqnarray}
where in the final step we used the fact that $\ba_{ji}\ba^{ik}=\delta_j^k$ so that 
\begin{eqnarray}
	(h_{ji}+\lambda F_{ji})\ba^{ik}=\delta_j^k \ , \quad 
	\ba^{ki}(h_{ij}+\lambda F_{ij})=\delta^k_j 
\end{eqnarray}
that when we combine together we obtain an important result 	
\begin{equation}
h_{ji}(
	\ba^{ik}-\ba^{ki})=-\lambda F_{ji}(\ba^{ik}+\ba^{ki}) \ . 
	\nonumber \\
\end{equation}
Before we proceed further we return to the definition of $\mR$ that can be written as
\begin{eqnarray}
	\mR=
	\tmH_i \bA^{ij}\tmH_j \ , \nonumber \\
\end{eqnarray}	
where the symmetric matrix $\bA^{ij}=\bA^{ji}$ is defined as
\begin{equation}
	\bA^{ij}=\frac{1}{\lambda^2}
	\pi^i\pi^j+\frac{1}{2}(\ba^{ij}+\ba^{ji})V^2\det \ba  \ . 
\end{equation}
Collecting all these results together we can write (\ref{pbbM}) in compact form 
\begin{eqnarray}\label{pbbMfinal}
	&&\pb{\bM(X),\bM(Y)}
=4\int d^9\bx(X\partial_iY-Y\partial_iX)\bA^{ij}\tmH_j  \ . 
	\nonumber \\ 
\end{eqnarray}
As the second step we  calculate following expression 
\begin{eqnarray}\label{pbbRM}
&&	\pb{\mathbf{R}(X),\bM(Y)}+\pb{\bM(X),\mathbf{R}(Y)}=
\nonumber \\
&&=2\int d^9\bx d^9\by X\tmH_i \bA^{ij}(\bx)\pb{\tmH_j(\bx),
	\mM(\by)}Y(\by)-\nonumber \\
&&-2\int d^9\bx d^9\by Y\tmH_i \bA^{ij}(\bx)\pb{\tmH_j(\bx),
	\mM(\by)}X(\by)+\nonumber \\
&&+\int d^9\bx d^9\by X \tmH_i\tmH_j(\bx)
\pb{\bA^{ij}(\bx),\mM(\by)}Y(\by)-\nonumber \\
&&-\int d^9\bx d^9\by
Y\tmH_i\tmH_j(\bx)\pb{\bA^{ij}(\bx),\mM(\by)}X(\by) \ ,
\nonumber \\		
\end{eqnarray}
where we again introduced smeared form of $\mR$ defined as
\begin{equation}
		 \mathbf{R}(X)\equiv
	\int d^9\bx X(\bx)\mR(\bx) \ .  \nonumber \\
	\end{equation}
First of all we consider formulas on the first two lines in 
(\ref{pbbRM}) where explicit calculation gives 
\begin{eqnarray}\label{pbbRM1}
&&2\int d^9\bx d^9\by X\tmH_i \bA^{ij}(\bx)\pb{\tmH_j(\bx),
	\mM(\by)}Y(\by)-\nonumber \\
&&-2\int d^9\bx d^9\by Y\tmH_i \bA^{ij}(\bx)\pb{\tmH_j(\bx),
	\mM(\by)}X(\by)=\nonumber \\
&&=4\int d^9\bx (X\partial_mY-Y\partial_mX)\bA^{mi}\tmH_i\mM \ . 
\nonumber \\
\end{eqnarray}
On the other hand  calculations on the last two lines in (\ref{pbbRM}) is much
more involved
\begin{eqnarray}\label{pbbRM2}
&&\int d^9\bx d^9\by X \tmH_i\tmH_j(\bx)
\pb{\bA^{ij}(\bx),\mM(\by)}Y(\by)-\nonumber \\
&&-\int d^9\bx d^9\by
Y\tmH_i\tmH_j(\bx)\pb{\bA^{ij}(\bx),\mM(\by)}X(\by)= 
\nonumber \\		
&&=2\int d^9\bx d^9\by Y\mH_i^G h^{ij}(\bx)
\pb{\mH^G_j(\bx),\det \ba \frac{1}{2}(\ba^{kl}+\ba^{lk})(\by)}V^2 X \tmH_k\tmH_l(\by)-\nonumber \\
&&-2\int d^9\bx d^9\by X\mH_i^G h^{ij}(\bx)
\pb{\mH^G_j(\bx),\det \ba \frac{1}{2}(\ba^{kl}+\ba^{lk})(\by)}V^2 Y \tmH_k\tmH_l(
\by)-\nonumber \\
&&-\int d^9\bx d^9\by X\tmH_i\tmH_j(\bx)\pb{(\frac{1}{\lambda^2}\pi^i\pi^j+V^2\det \ba \ba^{ij}_S)(
	\bx),
	(\frac{1}{\lambda^2}\pi^k h_{kl}\pi^l+V^2\det \ba)(\by)}Y(\by)+\nonumber \\
&&+\int d^9\bx d^9\by Y\tmH_i\tmH_j(\bx)\pb{(\frac{1}{\lambda^2}\pi^i\pi^j+V^2\det \ba \ba^{ij}_S)(
\bx),
(\frac{1}{\lambda^2}\pi^k h_{kl}\pi^l+V^2\det \ba)(
\by)}X(\by)=\nonumber \\
&&=-4\int d^9\bx (X\partial_mY-Y\partial_mX)\ba^{mn}_S\tmH_nV^2\det \ba\ba^{kl}
\tmH_k\tmH_l \nonumber \\
&&+2\int d^9\bx (X\partial_mY-Y\partial_mX)(\ba^{jm}+\ba^{mj})\tmH_j\tmH_i\ba^{ik}\tmH_kV^2\det \ba=0 \ , 
\nonumber \\
\end{eqnarray}
where we again used 
\begin{equation}
	h_{lk}\ba^{km}-\ba^{mk}h_{kl}=
	-\lambda F_{lk}(\ba^{km}+\ba^{mk}) \ , 
\end{equation}
and where $\ba^{ij}_S=\frac{1}{2}(\ba^{ij}+\ba^{ji})$. 
Collecting (\ref{pbbRM1}) and (\ref{pbbRM2}) together we obtain final result
\begin{eqnarray}\label{pbbRMfinal}
	\pb{\mathbf{R}(X),\bM(Y)}+\pb{\bM(X),\mathbf{R}(Y)}=4\int d^9\bx (X\partial_mY-Y\partial_m X)
	\bA^{mn}\tmH_n\mM \ . \nonumber \\	 
\end{eqnarray}
Finally we proceed to the calculation of Poisson bracket between smeared forms of $\mR$ 
%
\begin{eqnarray}\label{bRXY}
&&	\pb{\mathbf{R}(X),\mathbf{R}(Y)}=4\int d^9\bx d^9\by
	\tmH_i\bA^{ij}(\bx)\pb{\tmH_j(\bx),\tmH_k(\by)}
	\bA^{kl}\tmH_l(\by)+
	\nonumber \\
&&	+2\int d^9\bx d^9\by X\tmH_i\bA^{ij}(\bx)
	\pb{\tmH_j(\bx),\bA^{kl}(\by)}\tmH_k\tmH_l(\by)+
	\nonumber \\
&&	+2\int d^9\bx d^9\by
	X\tmH_i\tmH_j(\bx)\pb{\bA^{ij}(\bx),\tmH_k(\by)}
	\bA^{kl}\tmH_l(\by)+\nonumber \\
&&	+\int d^9\bx d^9\by
	\tmH_i\tmH_j(\bx)\pb{\bA^{ij}(\bx),
		\bA^{kl}(\by)}\tmH_k\tmH_l(\by) \ .
	\nonumber \\
\end{eqnarray}
First three terms in (\ref{bRXY}) can be easily calculated using
\begin{eqnarray}\label{bRXYh}
	\pb{\int d^9\bx X^i\tmH_i,
		\int d^9\by Y^j\tmH_j}=
	\int d^9\bx (X^m\partial_m Y^j-Y^m\partial_m X^j)\tmH_j \ , 
	\nonumber \\		
	\pb{\int d^9\bx X^i\tmH_i,\bA^{ij}}=
	-2\partial_m(X^m \bA^{ij})
	+\partial_m X^i\bA^{mj}+\bA^{im}\partial_m X^j \ .
	\nonumber \\	
\end{eqnarray}
In case of the last term in (\ref{bRXY}) the situation is more involved. For that reason  we define $\bA(Y_{ij})$ as
 \begin{equation}
	\bA(Y_{ij})=
	\int d^9\bx Y_{ij}\bA^{ij} \ , 
\end{equation}
where $Y_{ij}$ is arbitrary tensor function. Then we obtain 
\begin{eqnarray}\label{pbAXY}
&&	\pb{\bA(X_{ij}),\bA(Y_{kl})}=
2\int d^9\bx [(X_{ij}\pi^i\partial_m Y_{kl}-Y_{ij}\pi^i
	\partial_m X_{kl})(\ba^{jm}-\ba^{mj})V^2\det \ba \ba^{kl}+
	\nonumber \\
&&	+2\int d^9\bx [(X_{ij}\partial_m Y_{kl}-Y_{ij}\partial_m X_{kl})\pi^i(\ba^{mk}\ba^{lj}-
	\ba^{km}\ba^{jl})]V^2\det \ba \ .  \nonumber \\		
\end{eqnarray}
Collecting (\ref{bRXYh}) together with (\ref{pbAXY}) we obtain that (\ref{bRXY}) is equal to
\begin{eqnarray}\label{pbbRXY}
&&	\pb{\mathbf{R}(X),\mathbf{R}(Y)}=
	4
	\int d^9\bx (X\partial_mY-Y\partial_mX)\bA^{mn}\tmH_n \mR \ .\nonumber \\
\end{eqnarray}
Now we have all ingredient for calculations of Poisson brackets
between $\bH(X),\bH(Y)$ where $\bH(X)$ is defined as

\begin{equation}
\bH(X)=\int d^9\bx XH(\bx) \ , \quad 
H=
	\sqrt{\frac{\mM}{2}+\sqrt{\frac{1}{4}\mM^2-\mR}} \ 
\end{equation}	
and hence	
\begin{eqnarray}\label{pbHXY}
&&	\pb{\bH(X),\bH(Y)}
=\nonumber \\
&&	=\frac{1}{16}\int d^9\bx d^9
\by
\frac{X(\bx)}{\sqrt{\frac{1}{2}\mM+\sqrt{\frac{1}{4}\mM^2-\mR}}}
\pb{\mM(\bx),\mM(\by)}\frac{Y(\by)}{\sqrt{\frac{1}{2}\mM+\sqrt{\frac{1}{4}\mM^2-\mR}}}+
\nonumber \\
&&+	\frac{1}{16}\int d^9\bx d^9
\by
\frac{X(\bx)}{\sqrt{\frac{1}{2}\mM+\sqrt{\frac{1}{4}\mM^2-\mR}}}\times\nonumber \\
&&\pb{\mM(\bx),(\frac{1}{4}\mM^2-\mR)(\by)}\frac{Y(\by)}{\sqrt{\frac{1}{2}\mM+\sqrt{\frac{1}{4}\mM^2-\mR}}\sqrt{\frac{1}{4}\mM^2-\mR}}+
\nonumber \\
&&+	\frac{1}{16}\int d^9\bx d^9
\by
\frac{X(\bx)}{\sqrt{\frac{1}{2}\mM+\sqrt{\frac{1}{4}\mM^2-\mR}}\sqrt{\frac{1}{4}\mM^2-\mR}}\times\nonumber \\
&&\pb{(\frac{1}{4}\mM^2-\mR)(\bx),\mM(\by)}\frac{Y(\by)}{\sqrt{\frac{1}{2}\mM+\sqrt{\frac{1}{4}\mM^2-\mR}}}+
\nonumber \\
&&+	\frac{1}{16}\int d^9\bx d^9
\by
\frac{X(\bx)}{\sqrt{\frac{1}{2}\mM+\sqrt{\frac{1}{4}\mM^2-\mR}}\sqrt{\frac{1}{4}\mM^2-\mR}}\times\nonumber \\
&&\pb{(\frac{1}{4}\mM^2-\mR)(\bx),(\frac{1}{4}\mM^2-\mR)(\by)}\frac{Y(\by)}{\sqrt{\frac{1}{2}\mM+\sqrt{\frac{1}{4}\mM^2-\mR}}\sqrt{\frac{1}{4}\mM^2-\mR}}=0\nonumber \\
	\end{eqnarray}
using Poisson brackets (\ref{pbbMfinal}),(\ref{pbbRMfinal}) and (\ref{pbbRXY}).
In other words we got an important result 
\begin{equation}
	\pb{H(\bx),H(\by)}=0
\end{equation}
which holds even in case of finite tachyon field. Finally using this result we obtain 
\begin{eqnarray}\label{mGbxby}
	\pb{\mG(\bx),\mG(\by)}=\frac{1}{\sqrt{\lambda}}\pb{p_T(\bx),H(\by)}+
	\frac{1}{\sqrt{\lambda}}\pb{H(\bx),p_T(\by)}+\pb{H(\bx),H(\by)}=0   \nonumber \\
\end{eqnarray}
due the the fact that Poisson bracket $\pb{p_T(\bx),H(\by)}$ is ultralocal. 

Formulas (\ref{pbHXY}) and (\ref{mGbxby}) are  one of the most important
results presented in this paper. Explicitly they  say that deparametrized theory 
of space-time filling non-BPS D9-brane with any value of tachyon 
has vanishing Poisson brackets between new Hamiltonian constraints. This fact
has an important consequence for the construction of Dirac variables as
we will see in the next section.

Finally we will calculate Poisson bracket between $\bC_S(X^i)$ and $\mG$. Using
\begin{equation}
	\pb{\bC_S(X^i),H}=-X^m\partial_m H-\partial_m X^m H
\end{equation}
we obtain 
\begin{equation}
	\pb{\bC_S(X^i),\mG}=-X^m\partial_m\mG-\partial_m X^m\mG \ 
\end{equation}
that shows that $\mG$ transforms as tensor density. Finally note that it is easy to see that Poisson bracket between smeared form of spatial diffeomorphism constraints is equal to 
\begin{equation}
\pb{\bC_S(X^i),\bC_S(Y^j)}=\bC_S(X^j\partial_j Y^i-Y^j\partial_j X^i) \ .
\end{equation}
In other words we derived that  $\mG\approx 0,\mC_i\approx 0$ are first class constraints and that Poisson bracket between 
Hamiltonian constraints $\mG(\bx)\approx 0$ vanishes and hence we have all ingredients for construction of Dirac observables
which will be performed in next sections. 
\section{Deparametrization of General Relativity Coupled to Tachyon-Naive Treatment}\label{third}
We showed in \cite{Kluson:2024rnv}, following seminal work \cite{Thiemann:2006up}, 
 that in the case of  asymptotic large tachyon field when we neglect terms proportional to $V^2$ in $\mG$  we can define Dirac observable in very natural way. Explicitly, let us   define $\bH_\tau$ as
 \cite{Thiemann:2006up}
\begin{equation}
	\bH_\tau=\int d^9\bx[\tau-\sqrt{\lambda}T(\bx)]H(\bx) \ ,
\end{equation}
where $\tau$ has physical dimension of length which is appropriate for time variable.
Further, let $f$ is spatial diffeomorphism invariant quantity that does not depend on $T$ and $p_T$. 
Then let us define 
 $O_f(\tau)$ by following prescription
\begin{equation}\label{defOf}
	O_f(\tau)=\sum_{n=0}^\infty
	\frac{1}{n!} \pb{f,\bH_\tau}_{(n)} \ , 
\end{equation}
where multiple Poisson brackets are defined as 
\begin{eqnarray}
&&	\pb{f,\bH_\tau}_{(0)}=f \ , 
	\quad \pb{f,\bH_\tau}_{(1)}=\pb{\pb{f,\bH_\tau},\bH_\tau}_{(0)}=
	\pb{f,\bH_\tau} \ , \nonumber \\
&&	\pb{f,\bH_\tau}_{(n+1)}=\pb{\pb{f,\bH_\tau}_{(n)},\bH_\tau} \ . 
\end{eqnarray}
Then it can be shown that evolution with respect to the parameter $\tau$ has the form 

\begin{eqnarray}\label{eqQ}
&&	\frac{d}{d\tau}O_f(\tau)=\pb{O_f,\bH} \ , \quad \bH\equiv \int d^9\bx H 
\end{eqnarray}
which has the form of the Hamiltonian equation that expresses true evolution with respect to the time parameter $\tau$ generated by Hamiltonian $\bH$.
However the crucial question is whether $	O_f(\tau)$ defined above is true Dirac observable which 
means that it Poisson commutes with all first class constraints $\mC_i\approx 0 , \mG\approx 0$. In fact it is easy to demonstrate that such defined observable cannot be Dirac one simply from the fact that there is non-zero contribution from  Poisson bracket between $\mG(\bx)$ and $\bH_\tau$ 
that follows from explicit dependence of $H$ on $T$. In more details, let us calculate Poisson bracket between $p_T$ and $H$ 
\begin{eqnarray}
&&	\pb{\frac{1}{\sqrt{\lambda}}p_T(\bx),H(\by)}
	=\frac{1}{\sqrt{\lambda}}\frac{V\frac{dV}{dT}(\bx)}{H(\bx)}\det \ba (\bx)\delta(\bx-\by)
	+\nonumber \\
&&	+\frac{1}{2\sqrt{\lambda}}\frac{1}{H(\bx)}\frac{1}{2\sqrt{\frac{\mM^2}{4}-\mR}}V\frac{dV}{dT}\mM\det \ba
	\delta(\bx-\by)+\nonumber \\	
&&	+\frac{1}{2\sqrt{\lambda}}\frac{1}{H(\bx)}\frac{1}{\sqrt{\frac{\mM^2}{4}-\mR}}
	\frac{dV}{dT}V\det \ba \ba^{ij}\tmH_i\tmH_j\delta(\bx-\by) \  \nonumber \\
\end{eqnarray}
which is non-zero for finite $T$. Then it is easy to see that
\begin{equation}
	\pb{O_f(\tau),\bG(M)}\neq 0
\end{equation}
and hence $O_f(\tau)$ as was defined above is not Dirac observable. 
%
This is a consequence of the explicit dependence of $H$ on the tachyon field.
On the other hand  we can still define
Dirac observables when we proceed in the similar way as in the case of relational observables. 

\section{Gauge Invariant Observables for General $T$}
\label{fourth}
Let us outline general procedure of construction of relational  observables when  we mainly follow excellent papers by Dittrich \cite{Dittrich:2004cb,Dittrich:2005kc}.
Recall that we have ten first class constraints $\mC_i\approx 0, \mG\approx 0$ 
so that, following 
\cite{Dittrich:2004cb,Dittrich:2005kc}, we should consider  their smeared form
\begin{equation}
\bC(\Lambda)\equiv\int d^9\bx (\Lambda^G\mG+\Lambda^i\mC_i)
\equiv \int d^9\bx \Lambda^K\tilde{\mC}_K \ , K=0,1,\dots,9 \ , 
\end{equation}
where $\Lambda^K$ are ten smeared functions. 
The gauge transformations generated by the function $\bC(\Lambda)$ has the form 
\begin{equation}
	\alpha_{\bC(\Lambda)}(f(\bx))=
	\sum_{r=0}^\infty
	\frac{1}{r!}\pb{f(\bx),\bC(\Lambda)}_r \ , 
\end{equation}
where
\begin{eqnarray}
&&	\pb{f,\bC(\Lambda)}_0=f \ , \quad 
	\pb{f,\bC(\Lambda)}_1=\pb{f,\bC(\Lambda)} \ , \nonumber \\ 
&&	\pb{f,\bC(\Lambda)}_{r+1}=\pb{\pb{f,\bC(\Lambda)}_r,\bC(
		\Lambda)} \ . 
\end{eqnarray}
Observables, which are invariant under these gauge transformations, are called Dirac observables. 
Partial observables  are phase space functions which are not invariant under gauge transformations. It is convenient to split these partial variables into clocks variables $T_K(\bx)$ where $K$ label first class constraints and remaining ones that we denote as $f$. Note that generally we need as many partial variables $T_K$ as there are first class constraints.

%

The complete observable $F_{[f;T]}(\tau)(P)$ associated to the partial observable $f$ and the clock variables $T^K(\bx)$ will generally depend on infinite many parameters $\tau^K(\bx)$. It gives value of 
the phase space function at the phase space point $Q$ in  the gauge orbit through the phase space point $P$
when the clock variables give the values
$T^K(\bx)(Q)=\tau^K(\bx)$ for all $\bx\in \Sigma$ and for all $K$.

Generally the calculation of complete observable is very difficult. First of all we find point $Q$ on the gauge orbit through the point $P$ at which $[T^K(\bx)](Q)=\tau^K(\bx)$. In other words we calculate the flow
\begin{equation}\label{lambdaT}
	[\alpha_{\bC(\Lambda)}(T(\bx))](P)=\sum_{r=0}^\infty\frac{1}{r!}\pb{T(\bx),\bC(\Lambda)}_r
\end{equation}
and find function $\beta(\bx)$ such that
\begin{equation}\label{alphaT}
	[\alpha_{\bC(\Lambda)}(T^K(\bx))]_{\Lambda\rightarrow \beta(P)}(P)\approx\tau^K(\bx)
\end{equation}
for all $\bx \in \Sigma$ where $\approx $ means that this equation holds on the constraint surface only.
It is important to stress the difference between smeared function $\Lambda$ which depends on $\bx$ only, and
$\beta(P)$ that generally depends on the phase space point $P$. For that reason in the equations above we should firstly calculate all Poisson brackets and only  then to replace $\Lambda$ with $\beta$.  The value of complete observable is then given by expression \cite{Dittrich:2004cb,Dittrich:2005kc}
\begin{equation}
	F_{[f;T]}(\tau,P)=
	[\alpha_{\bC(\Lambda)}(f)]_{\Lambda\rightarrow \beta(P)}(P) \ . 
\end{equation}
As was shown in \cite{Dittrich:2004cb,Dittrich:2005kc}   it is very difficult to find explicit form of complete observable due to the fact that
(\ref{alphaT}) leads to functional differential equations. 

In case of the constraint $\mG\approx 0$ the situation is simpler since
\begin{equation}
	\pb{T(\bx),\mG(\by)}_0=T(\bx) \ , 
	\quad 
	\pb{T(\by),\mG(\bx)}_1=\frac{1}{\sqrt{\lambda}}\delta(\bx-\by)
\end{equation}
while we also have
\begin{equation}
	\pb{T(\bx),\bC(\Lambda^i)}=\Lambda^i\partial_iT 
\end{equation}
and hence
\begin{equation}\label{eqT}
	[\alpha_{\bC(\Lambda)}(T_K(\bx))](P)=\sum_{r=0}^\infty\frac{1}{r!}\pb{T_K(\bx),\bC(\Lambda)}_r=T(\bx)+\frac{1}{\sqrt{\lambda}}\Lambda^G(\bx)+
	\Lambda^i\partial_iT  \ . 
\end{equation}
It is important to stress that in case of world-volume of space-time filling non-BPS D9-brane there are no additional modes that could serve as spatial "clock" variables $T^i(\bx)$.
In other words the general procedure of construction 
of Dirac variables is not directly applicable in case of space-time filling non-BPS D9-brane.
For that reason we slightly generalize procedure suggested in \cite{Thiemann:2006up}. Explicitly, 
let us define following observable $F_{[f;T]}$ 
\begin{eqnarray}\label{FfTconst}
&&	F_{[f;T]}=
	\sum_{r=0}^\infty
	\frac{1}{r!}\int_\Sigma d\bx_1\dots d\bx_r
		(\tau-\sqrt{\lambda}T(\bx_1))\times\dots
	(\tau-\sqrt{\lambda}T(\bx_r))\times \nonumber \\
&&\times 	\pb{f,\mG(\bx)}_r 
\ . 
\end{eqnarray}
where $f$ is phase space function that is invariant under spatial diffeomorphism so that it has the form of integral of space density over spatial section. Further we should also stress that parameter 
$\tau$ that appears in (\ref{FfTconst}) does not depend on $\bx$.

Let us now calculate derivative of $F_{[f;T]}$ with respect to $\tau$
\begin{eqnarray}\label{eqF}
&&	\frac{d F_{[f;T]}}{d\tau}
=\sum_{r=0}^\infty \int_\Sigma d^9\bx_1\dots
d\bx_n \frac{1}{(r-1)!}(\tau-\sqrt{\lambda}T(\bx_2))
\times \dots\times (\tau-\sqrt{\lambda}T(\bx_{(r-1)}))
\times \nonumber \\
&&\times\pb{\dots\{\pb{f,\mG(\bx_1)},\mG(\bx_2)\},\dots,
	\mG(\bx_r)}=
	F_{[\pb{f,\bG(1)};T]} \ ,
	\nonumber \\
\end{eqnarray}
where
\begin{equation}
	\bG[1]=\int d^9\bx \mG(\bx) \ , 
\end{equation}
and 
where we used  Jacobi identity 
\begin{eqnarray}
	\pb{\mG(\bx_1),\pb{\mG(\bx_2),f}}+
	\pb{\mG(\bx_2),\pb{f,\mG(\bx_1)}}+
	\pb{f,\pb{\mG(\bx_1),\mG(\bx_2)}}=0
	\nonumber \
\end{eqnarray}
that using the fact that $\pb{\mG(\bx_1),\mG(\bx_2)}=
\pb{\mH(\bx_1),\mH(\bx_2)}=0$ implies
\begin{equation}
	\pb{\mG(\bx_1),\pb{\mG(\bx_2),f}}=\pb{\mG(\bx_2),
			\pb{\mG(\bx_1),f}} \ . 
\end{equation}
We see that (\ref{eqF}) does not have the form of the Poisson bracket
between $F_{[f;T]}$ and Hamiltonian function  which was the case when 
Hamiltonian constraint does not depend on $T$ explicitly. On the other hand
it is clear that in the limit $T\rightarrow \infty$ when we can neglect
$V^2$ in $H$ we find that $p_T$ Poisson commutes with $H$ so that 

\begin{eqnarray}\label{eqF}
	&&	\lim_{T\rightarrow \infty}\frac{d F_{[f;T]}}{d\tau}
	=\sum_{r=0}^\infty \int_\Sigma d^9\bx_1\dots
	d\bx_n \frac{1}{(r-1)!}(\tau-\sqrt{\lambda}T(\bx_2))
	\times \dots\times (\tau-\sqrt{\lambda}T(\bx_{(r-1)}))
	\times \nonumber \\
	&&\times\pb{\dots\{\pb{f,H(\bx_1)},H(\bx_2)\},\dots,
		H(\bx_r)}=
\pb{F_{[f;T]},\bH} \nonumber \\
\end{eqnarray}
so that we reproduce result derived in \cite{Kluson:2024rnv} which is nice
consistency check. 
Let us also show that $F_{[f;T]}$ has vanishing Poisson bracket with $\bG(M)$. 
In order to simplify notation let us define function 
 $F_r(\bx_1,\dots,\bx_r)$ as 
\begin{eqnarray}
&&	F_r(\bx_1,\dots,\bx_r)=\pb{\dots\pb{f,\mG(\bx_1)},\mG(\bx_2)
		\},\dots\},\mG(\bx_r)}\equiv
		\nonumber \\
&&\equiv 		\pb{F_{(r-1)}(\bx_1,\dots,\bx_{r-1}),
		\mG(
		\bx_r)} \ .  \nonumber \\
\end{eqnarray}
Then we have
\begin{eqnarray}
&&\int d^9\bx_1\dots d^9\bx_r	\pb{(\tau-\sqrt{\lambda}T(\bx_1))\times \dots\times 
		(\tau-\sqrt{\lambda}T(\bx_r))
			F_r(\bx_1,\dots\bx_r),\bG(M)}=
\nonumber \\
&&-r\int d^9\bx_1\dots d^9\bx_{(r-1)}
(\tau-\sqrt{\lambda}T(
\bx_1))
\times \nonumber \\
&& \dots \times (
\tau-
\sqrt{\lambda}T(\bx_{(r-1)}))\pb{F_{r-1}(\bx_1,
	\dots,\bx_{r-1}),\bG(M)}+\nonumber \\	
&&+\int d^9\bx_1\dots d^9\bx_r
(\tau-\sqrt{\lambda}T(\bx_1))
\times \dots (\tau-\sqrt{\lambda}T(\bx_r))
\pb{F_r(\bx_1,\dots,\bx_r),\bG(M)} \nonumber \\		
\end{eqnarray}
where we again used  Jacobi identity 
\begin{eqnarray}\label{Jacgen}
	\pb{\pb{X,\mG(\bx)},\bG(M)}+\pb{\pb{\mG(\bx),\bG(M)},X}+\pb{\pb{\bG(M),X},\mG(\bx)}=0
	\nonumber \\
\end{eqnarray}
that holds for any phase space function $X$. Since 
 $\pb{\mG(\bx),\bG(M)}=0$ we see that (\ref{Jacgen}) implies 
\begin{equation}
	\pb{\pb{X,\mG(\bx)},\bG(M)}=\pb{\pb{X,\bG(M)},\mG(\bx)} \ . 
\end{equation}
Using these results we obtain 
\begin{eqnarray}
&&	\pb{F_{[f;T]},\bG(M)}=
\pb{f,\bG(M)}+\nonumber \\
&&+\sum_{r=1}^\infty
\frac{1}{r!}\int_\Sigma d\bx_1\dots \bx_r
\pb{(\tau-\sqrt{\lambda}T(\bx_1))
	(\tau-\sqrt{\lambda}T(\bx_r))
	F_r(\bx_1,\dots\bx_r),\bG(M)}=
\nonumber \\
&&\pb{f,\bG(M)}-\pb{f,\bG(M)}
-\sum_{r=2}^\infty 
\frac{1}{(r-1)!}
\int d^9\bx_1\dots d^9\bx_{(r-1)}
(\tau-\sqrt{\lambda}T(
\bx_1))
\times \nonumber \\
&&\dots \times (
\tau-
\sqrt{\lambda}T(\bx_{(r-1)}))\pb{F_{r-1}(\bx_1,
	\dots,\bx_{(r-1)}),\bG(M)}+\nonumber \\	
&&+\sum_{r=1}^\infty\int d^9\bx_1\dots d^9\bx_r
(\tau-\sqrt{\lambda}T(\bx_1))
\times \nonumber \\
&& \dots \times (\tau-\sqrt{\lambda}T(\bx_r))
\pb{F_r(\bx_1,\dots,\bx_r),\bG(M)} )=0\nonumber \\	
\end{eqnarray}
which is desired result. 
\subsection{Spatial Diffeomorphism Constraint}
Finally we will check that $F_{[f;T]}$ Poisson commutes with $\bC_S(M^i)$. Since $f$ is spatially invariant function by definition we have 
\begin{equation}\label{spatO}
	\pb{\bC_S(M^i),f}=0 \ .
\end{equation}
Further, since $\mG(\bx)$ is tensor density we immediately obtain 
\begin{equation}
	\pb{\bC(M^i),\mG(\bx)}=-\partial_i M^i\mG(\bx)-M^i\partial_i\mG(\bx) \ . 
\end{equation}
Let us further calculate Poisson bracket between the 
function $F_r(\bx_1,\dots,\bx_r)$ and $\bC(M^i)$
\begin{eqnarray}
&&	\pb{F_r(\bx_1,\dots,\bx_r),\bC(M^i)}=
	\pb{\pb{F_{r-1}(\bx_1,\dots,\bx_{(r-1)})},\mG(\bx_r),\bC(M^i)}=\nonumber \\
&&	=-\pb{\pb{\mG(\bx_r),\bG(M)},F_{r-1}(\bx_1,\dots,\bx_{(r-1)})}
-\pb{\pb{\bC(M^i),F_{r-1}(\bx_1,\dots,\bx_{(r-1)})},\mG(\bx_r)}=\nonumber \\
&& =	M^i(\bx_r)\frac{\partial}{\partial x^i_r}F_r(\bx_1,\dots,\bx_r)
	+\frac{\partial M^i(\bx_r)}{\partial x^i_r}F_r(\bx_1,\dots,\bx_r)
	+\nonumber \\
&&+M^i(\bx_{r-1})\frac{\partial}{\partial x^i_{r-1}}F_r(\bx_1,\dots,\bx_r)+\frac{\partial M^i(\bx_{r-1})}{\partial x^i_{r-1}}F_r(\bx_1,\dots,\bx_r)+\dots \nonumber \\
&&+M^i(\bx_1)\frac{\partial}{\partial x^i_1}F_r(\bx_1,\dots,\bx_r)
+\frac{\partial M^i(\bx_1)}{\partial x^i_1}F_r(\bx_1,\dots,\bx_r) \ , 
\nonumber \\
		\end{eqnarray}
where we used the fact that 
\begin{eqnarray}
	\pb{\partial_{x^i}\mG(\bx),X(\by)}=\partial_{x^i}\pb{\mG(\bx),X(\by)}
\end{eqnarray}
that holds for any phase space function $X$ that does not depend on $\bx$.
Then it is easy to see that
\begin{eqnarray}
&&\pb{\int d^9\bx_1\dots d^9\bx_r(\tau-T(\bx_1))
	\times (\tau-T(\bx_r))F_r(\bx_1,\dots,\bx_r),\bC(M^i)}=
	\nonumber \\
&&=\int d^9\bx_1 \dots d^9\bx_r (\sqrt{\lambda}M^i(\bx_1)\frac{T}{\partial x^i_1}(\tau-\sqrt{T}(\bx_2)\times \dots\times (\tau-\sqrt{\lambda}T(\bx_r))+
\dots \nonumber \\
&&M^i(\bx_r)\frac{\partial T}{\partial x^i_r}
(\tau-T(\bx_1))\dots (\tau-T(\bx_{(r-1)})))
F_r(\bx_1,\dots,\bx_r) +\nonumber \\
&&+\int d^9\bx_1\dots d^9\bx_r
(\tau-T(\bx_1))
\times (\tau-T(\bx_r))\times (\nonumber \\
|&& \times 	M^i(\bx_r)\frac{\partial}{\partial x^i_r}F_r(\bx_1,\dots,\bx_r)
+\frac{\partial M^i(\bx_r)}{\partial x^i_r}F_r(\bx_1,\dots,\bx_r)
+\nonumber \\
&&+M^i(\bx_{r-1})\frac{\partial}{\partial x^i_{r-1}}F_r(\bx_1,\dots,\bx_r)
+\frac{\partial M^i(\bx_{r-1})}{\partial x^i_{r-1}}F_r(\bx_1,\dots,\bx_r)+\dots \nonumber \\
&&+M^i(\bx_1)\frac{\partial}{\partial x^i_1}F_r(\bx_1,\dots,\bx_r)
+\frac{\partial M^i(\bx_1)}{\partial x^i_1}F_r(\bx_1,\dots,\bx_r))=0\nonumber \\
\end{eqnarray}	
and consequently
\begin{equation}
	\pb{F_{[f;T]},\bC_S(M^i)}=0 \ . 
\end{equation}
We see that the observable $F_{[f;T]}$ has vanishing Poisson brackets with all first class
constraints and it is true Dirac observable. This result shows that we can still define
these variables even in case of finite tachyon however now the time evolution of these variables
is more complicated than in case of asymptotic large tachyon. In fact, it is easy to see that
the observable $F_{[f;T]}$ reduces to observable introduced in 
\cite{Kluson:2024rnv} for large tachyon $T$. However the main message of this analysis is the fact that
 open string tachyon   is natural time variable even in case when we cannot neglect term proportional to $V^2$ in Hamiltonian constraint. We mean that this result is further support for  A. Sen's proposal
\cite{Sen:2023qya,Sen:2002qa}.

{\bf Data availability statement }

Data sharing not applicable to this article as no
datasets were generated or analyzed.

{\bf Acknowledgment:}

This work  is supported by the grant “Dualities and higher order derivatives” (GA23-06498S) from the Czech Science Foundation (GACR).

\end{document}